\documentclass[preprint,12pt,authoryear,harvard]{elsarticle}

\usepackage[authoryear]{natbib}

\usepackage{amssymb}
\usepackage{amsmath}

\journal{CMS}

\begin{document}
\newcommand{\etal}{{\em et al.}}{}
\newcommand{\fig}[1]{Fig.~\ref{#1}}
\newcommand{\tab}[1]{{Table ~(\ref{#1})}}
\newcommand{\eqn}[1]{{Eq.~(\ref{#1})}}
\newcommand{\AAA}{\AA\,}{}    
\newcommand{\eVA}{$\rm{eV/\rm{\AA}}$\,}{} 
\newcommand{\ww}{1.0}
\begin{frontmatter}



\title{Quantum Mechanical Simulations of Nanoindentation of Al Thin Film}


\author{Qing Peng, Xu Zhang and Gang Lu}

\address{ Department of Physics and Astronomy, California State
University Northridge, \\Northridge, CA, USA}

\begin{abstract}
QCDFT is a multiscale modeling approach that can simulate
multi-million atoms {\it effectively} via density functional theory
(DFT). The method is based on the framework of quasicontinuum (QC)
approach with DFT as its sole energetics formulation. The local QC
energy is calculated by DFT with Cauchy-Born hypothesis and the
nonlocal QC energy is determined by a self-consistent embedding
approach, which couples nonlocal QC atoms to the vertices of the
finite-elements at the local QC region. The QCDFT method is applied
to a nanoindentation study of an Al thin film in the presence and
absence of Mg impurities. The results show that the randomly
distributed Mg impurities can significantly increase the ideal and
yield strength of the Al thin film.
\end{abstract}


\end{frontmatter}





\section{Introduction}
The ability to perform quantum mechanical simulations of materials
properties over length scales that are relevant to experiments
represents a grand challenge in computational materials science. If
one could treat multi-millions or billions of electrons {\it
effectively} at micron scales, such first-principle quantum
simulations could revolutionize materials research and pave the way
to the computational design of advanced materials.

In this paper, we propose a multiscale approach that is based {\it
entirely} on density functional theory (DFT) and allows quantum
simulations at micron scale and beyond. The method, termed
QCDFT\citep{Peng2008}, combines the coarse graining idea of the
quasicontinuum (QC) approach and the coupling strategy of the
quantum mechanics/molecular mechanics (QM/MM) method, and represents
a major advance in the quantum simulation of materials properties.
It should be stated at the outset that QCDFT is {\it not} a
brute-force electronic structure method, but rather a multiscale
approach that can treat large systems - effectively up to billions
of electrons. Therefore, some of the electronic degrees of freedom
are reduced to continuum degrees of freedom in QCDFT. On the other
hand, although QCDFT utilizes the idea of QM/MM coupling, it does
not involve any classical/empirical potentials (or force fields) in
the formulation - the energy calculation of QCDFT is entirely based
on DFT. This is an important feature and advantage of QCDFT, which
qualifies it as a bona fide quantum mechanical simulation method.

Since QCDFT is formulated within the framework of the QC method, we
shall give a brief introduction to QC in Sec. 2.1 to set up the
stage of QCDFT. In Sec. 2.2, we briefly explain the local QC
calculations. In Sec. 2.3, we introduce a DFT-based QM/MM approach
that can treat the nonlocal QC region accurately and efficiently. In
Sec. 3, we apply QCDFT to the study of nanoindentation of an Al thin
film in the presence and absence of Mg impurities. We present the
nanoindentation results in Sec. 4 and finally our conclusions in
Sec. 5.

\section{QCDFT Methodology}

\subsection{Quasicontinuum Method}
The goal of the QC method is to model an atomistic system without
explicitly treating every atom in the problem
\citep{Tadmor1996,Shenoy1999}. This is achieved by replacing the
full set of $N$ atoms with a small subset of $N_r$ ``representative
atoms'' or {\it repatoms} ($N_r\ll N$) that approximate the total
energy through appropriate weighting. The energies of individual
repatoms are computed in two different ways depending on the
deformation in their immediate vicinity. Atoms experiencing large
variations in the deformation gradient on an atomic scale are
computed in the same way as in a standard atomistic method. In QC
these atoms are called {\em nonlocal} atoms. In contrast, the energy
of atoms experiencing a smooth deformation field on the atomic scale
is computed based on the deformation gradient ${\bf G}$ in their
vicinity as befitting a continuum model. These atoms are called {\em
local} atoms because their energy is based only on the deformation
gradient at the point where it is computed. In a classical system
where the energy is calculated based on classical/empirical
interatomic potentials, the total energy $E_{\rm tot}$ can be
written
\begin{equation}
\label{eq:energyqcoriginal} E_{\rm {tot}}^{\rm
{QC}}=\sum_{i=1}^{N^{{\rm nl}}}E_i({\bf R}) +\sum_{j=1}^{N^{{\rm
loc}}}n_jE_j^{{\rm loc}}({\bf G}).
\end{equation}
The total energy has been divided into two parts: an atomistic
region of $N^{\rm nl}$ nonlocal atoms and a continuum region of
$N^{{\rm loc}}$ local atoms ($N^{{\rm nl}}+N^{{\rm loc}}=N_r$). The
calculation in the nonlocal region is identical to that in atomistic
methods with the energy of the atom depending on the coordinates
${\bf R}$ of the surrounding repatoms. Rather than depending on the
positions of neighboring atoms, the energy of a local repatom
depends on the deformation gradients ${\bf G}$ characterizing the
finite strain around its position.

\subsection{Local QC calculation with DFT}
In the local QC region, a finite element mesh is constructed with
each repatom on the vertices of surrounding finite elements. The
energy and force of each local repatom can be obtained from the
strain energy density and the stress tensor of the finite elements
that share the same repatom. More specifically, according to the
Cauchy-Born rule, the deformation gradient ${\bf G}$ is the same for
a given finite element, therefore the local energy density
$\varepsilon$ and the stress tensor for each finite element can be
calculated as a perfect infinite crystal undergoing a uniform
deformation specified by ${\bf G}$. In other words, one could
perform a DFT-based energy/stress calculation for an infinite
crystal by using periodic boundary conditions with the primitive
lattice vectors of the deformed crystal, ${\mathbf{h}_i}$ given by
\begin{equation}
 \mathbf{h}_i=\mathbf{G} \, \mathbf{H}_i, \qquad \qquad \qquad i=1, 2, 3.
\label{eq:vector}
\end{equation}

\begin{figure}[htp]
\centering
\includegraphics[width=\ww\textwidth]{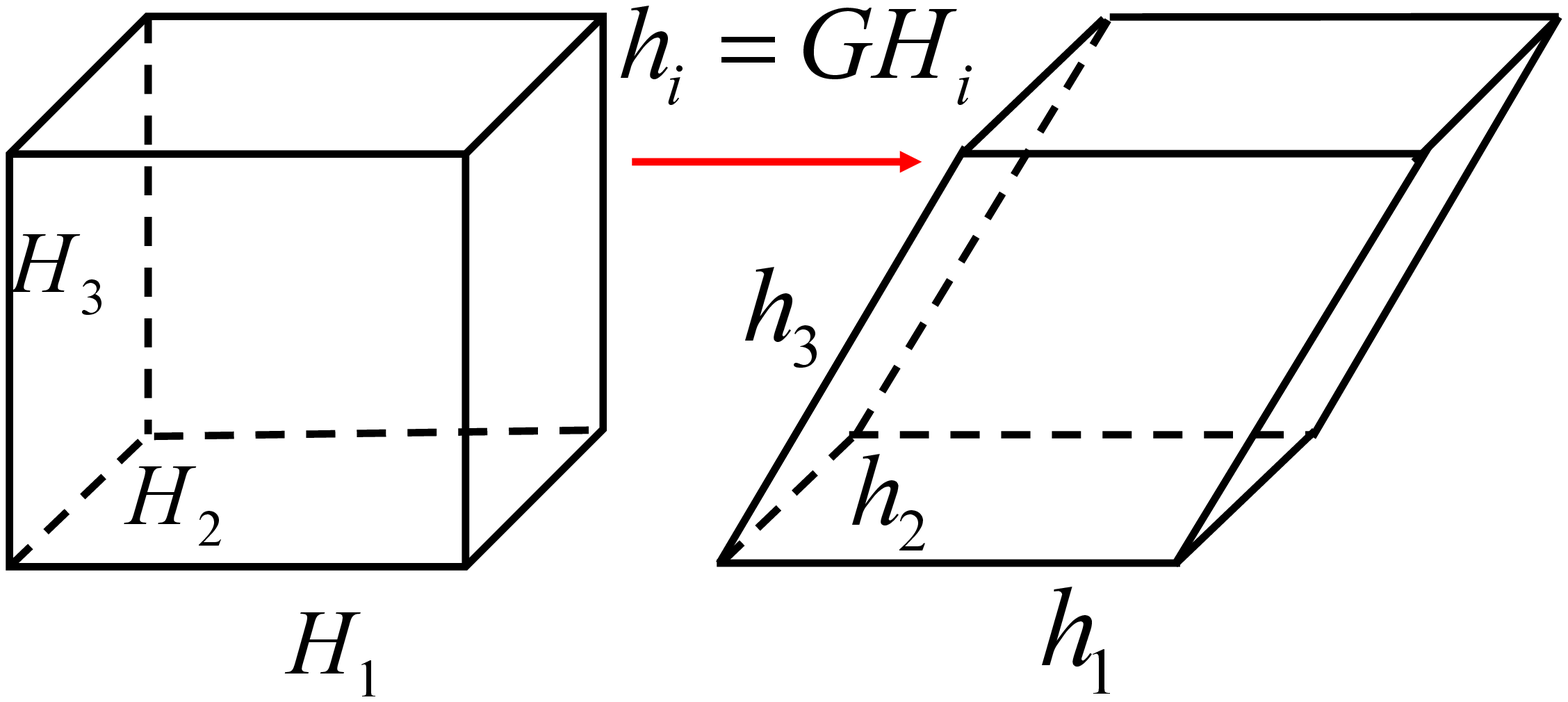}
 \caption{\label{fig:G} The deformed crystal derived from the perfect crystal via
 the deformation gradient ${\bf G}$.}
\end{figure}

Here $\mathbf{H}_i$ are the primitive lattice vectors of the perfect
undeformed crystal and $\Omega_0$ is the volume of the primitive
unit cell. The deformed crystal can be derived from the perfect
crystal via the deformation gradient ${\bf G}$ as shown in
\fig{fig:G}

For the deformation gradient ${\bf G_k}$ associated with the $k$th
element, a periodic DFT calculation can be performed to determine
the strain energy per unit cell $E^{\rm{DFT}}(\mathbf{G_k})$. The
Cauchy stress tensor can be defined as follows:
\begin{equation}
\sigma_{ab}=\frac{1}{\Omega}\sum_{\nu} \frac{\partial
E^{\rm{DFT}}(\mathbf{G}_k)}{\partial h_{a \nu} } h_{b\nu}
\end{equation}
with $\Omega$ being the volume of the deformed unit cell and
$h_{ij}$ denoting the component of the deformed lattice vector $h_j$
in Cartesian coordinate $i$.

Once the strain energy $E^{\rm{DFT}}(\mathbf{G}_k)$ is determined,
the energy contribution of the $j$th local repatom is given as
\begin{equation}
E_j^{{\rm loc}}(\{{\bf G}\})=\sum^{M_j}_{k=1} w_{jk}
E^{\rm{DFT}}(\mathbf{G}_k),
\end{equation}
where $M_j$ is the total number of finite elements shared by the
$j$th repatom, and $w_{jk}$ is the weight associated with the $k$th
finite element for the $j$th local repatom. The force on the $j$th
local repatom is defined as the gradient of the total energy with
respect to its coordinate $\textbf{R}_{j}^{{\rm loc}}$. In practice,
the nodal force on each finite element is calculated from the stress
tensor of the finite element by using the Principle of Virtual Work
\citep{FEMbook2000}. The force on the repatom is then obtained by
summing the nodal force contributions from each surrounding finite
elements.

\subsection{Nonlocal QC calculation with DFT}

\begin{figure}[htp]
\centering
\includegraphics[width=0.59\textwidth]{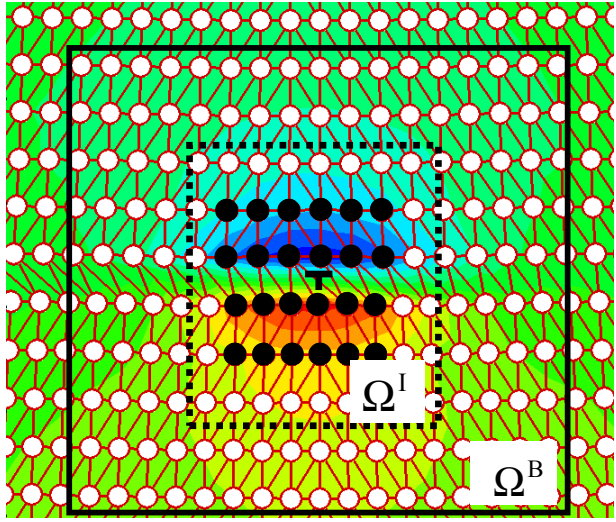}
 \caption{\label{fig:omega} The schematic diagram of domain
partition in QCDFT with a dislocation in Al lattice as an example.
The black and white spheres represent the nonlocal and buffer atoms,
respectively. The dotted box represents $\Omega^{\rm I}$ and the
solid box represents the periodic box $\Omega^{\rm B}$.  The volume
$\Omega^{\rm I}$ and $\Omega^{\rm B}$ is 2.8 \AAA and 8 \AAA beyond
the region I in $\pm x$ and $\pm y$ directions, respectively.}
\end{figure}

The nonlocal QC is modeled at the atomistic level with a QM/MM
approach. In a typical QM/MM calculation, the system is partitioned
into two domains: a QM region and an MM region. In QCDFT, the QM
atoms refer to the nonlocal repatoms and the MM atoms refer to the
buffer atoms which are the combination of both dummy atoms and local
repatoms in QC terminology. The so-called dummy atoms are in the
local region and are not independent degrees of freedom, but rather
slaves to the local repatoms. In other words, the position of a
dummy atom is determined by the finite element interpolation from
the relevant local repatom positions\citep{Tadmor1996,Shenoy1999}.
The dummy atoms provide the appropriate boundary conditions for
nonlocal DFT calculation while the energy of the dummy atoms is
still treated with the Cauchy-Born rule, consistent with their
status. The self-consistent embedding theory
\citep{Choly2005,Xu2007,Xu2008} is employed for the QM/MM
calculations. More specifically, both the energy of the nonlocal
atoms and the interaction energy between the nonlocal atoms and the
buffer atoms are calculated by DFT. To simply the notation, we
denote the nonlocal region as region I, and the buffer region as
region II, as shown in \fig{fig:omega}. Typically, the buffer region
consists of several atomic layers surrounding the nonlocal region.
We associate each buffer atom in region II with a valence electron
density ($\rho^{\rm at}$) and a pseudopotential; both of them are
constructed {\it a priori} and remain fixed during a QM/MM
simulation. \citep{Xu2007} The nonlocal energy $E^{\rm nl}$ as
defined in Eq. (1) can be expressed as
\begin{equation}
E^{\rm nl}={\rm min}_{\rho^{\rm I}}\{E_{\rm DFT}[\rho^{\rm
I};\textbf{R}^{\rm I}]+E_{\rm OF}^{\rm int}[\rho^{\rm I},\rho^{\rm
II};\textbf{R}^{\rm I},\textbf{R}^{\rm II}]\}.
\end{equation}
Here $\textbf{R}^{\rm I}$ and $\textbf{R}^{\rm II}$ denote atomic
coordinates in region I and II respectively. The charge density of
region I, $\rho^{\rm I}$, is the degree of freedom and is determined
self-consistently by minimizing the nonlocal energy functional. The
charge density of region II, $\rho^{\rm II}$, is defined as the
superposition of atomic-centered charge densities $\rho^{\rm at}$
via $\rho^{\rm II}(\textbf{r})=\sum_{i\in{\rm II}}\rho^{\rm
at}(\textbf{r}-\textbf{R}_{i})$, which only changes upon the
relaxation of region II ions. $E_{\rm OF}^{\rm int}$ is the
interaction energy between regions I and II computed by orbital-free
DFT (OFDFT) \citep{ofdftsurveywang,wang1992}.

A basic ansatz of the nonlocal energy functional (Eq. (5)) is that
$\rho^{\rm I}$ must be confined within a finite volume ($\Omega^{\rm
I}$) that is necessarily {\it larger} than region I but much smaller
than the entire QM/MM region. In addition, since some terms in the
formulation of Eq. (5) could be more efficiently computed in
reciprocal space \citep{Xu2007}, we also introduce a volume
$\Omega^{\rm B}$ over which the periodic boundary conditions are
applied. The periodic box $\Omega^{\rm B}$ should be large enough to
avoid the coupling errors induced by the implementation of periodic
boundary condition \citep{Xu2007}.

The interaction energy, $E_{\rm OF}^{\rm int}$, formulated by OFDFT
is defined as following:
\begin{equation}
\begin{split}
E_{\rm OF}^{\rm int}[\rho^{\rm I},\rho^{\rm II};\textbf{R}^{\rm
I},\textbf{R}^{\rm II}]&=E_{\rm OF}[\rho^{\rm tot};\textbf{R}^{\rm
tot}]
-E_{\rm OF}[\rho^{\rm I};\textbf{R}^{\rm I}]\\
&-E_{\rm OF}[\rho^{\rm II};\textbf{R}^{\rm II}],
\end{split}
\end{equation}
where $\textbf{R}^{\rm tot}\equiv\textbf{R}^{\rm
I}\bigcup\textbf{R}^{\rm II}$ and $\rho^{\rm tot}=\rho^{\rm
I}+\rho^{\rm II}$. In addition to its computational efficiency,
OFDFT allows Eq. (9) to be evaluated over $\Omega^{\rm I}$ rather
than over the entire QM/MM system as Eq. (9) appears to suggest
\citep{Choly2005,Xu2007}. This significant computational saving is
due to the cancellation in evaluating the first and second term of
Eq. (9), and it is rendered by the orbital-free nature of OFDFT and
the localization of $\rho^{\rm I}$. A single-particle embedding
potential $\mu_{\rm emb}(\textbf{r})$ can be defined as a functional
derivative of the interaction energy with respect to $\rho^{\rm I}$
\begin{equation}
\begin{split}
\mu_{\rm emb}(\textbf{r})\equiv\frac{\delta{E}_{\rm OF}^{\rm
int}[\rho^{\rm I}, \rho^{\rm II}; \textbf{R}^{\rm I},\textbf{R}^{\rm
II}]}{\delta\rho^{\rm I}},
\end{split}
\end{equation}
which represents the effective potential that region I electrons
feel due to the presence of region II; \citep{Choly2005,Xu2007} it
is through $\mu_{\rm emb}(\textbf{r})$ that the QM/MM coupling is
achieved quantum mechanically at the level of OFDFT. The embedding
potential provides rigorous boundary conditions for $\rho^{\rm I}$
and is updated self-consistently during the minimization of the
nonlocal energy functional.

Since $\rho^{\rm II}$ is a key quantity for the accurate calculation
of the interaction energy and the embedding potential, it is crucial
to construct an appropriate representation of $\rho^{\rm II}$. In
fact, the construction of an appropriate charge density distribution
in region II represents a common challenge to many QM/MM methods.
\citep{Lin2007} In this paper, we represent $\rho^{\rm II}$ as a
superposition of spherical atomic-like charge densities centered on
each ions in region II, which is a good approximation for metallic
systems. Ideally, the constructed $\rho^{\rm II}(\textbf{r})$ should
reproduce the bulk (or solid) charge density obtained by a DFT
calculation of the perfect lattice. That is to say, one needs to
determine $\rho^{\rm at}(r)$ by minimizing the function
\begin{equation}
\label{equ:sec2:024}
\begin{split}
\int_{\rm V_{\rm u}}[\rho^{\rm II}(\textbf{r})-\rho^{\rm
solid}(\textbf{r})]^{2}d\textbf{r}
\end{split}
\end{equation}
with $\rho^{\rm II}(\textbf{r})=\sum_{\mu}\rho^{\rm
at}(\textbf{r}-\textbf{R}_{\mu})$. Here $V_{\rm u}$ represents the
volume of the unit cell, and $\rho^{\rm solid}$ is the solid charge
density obtained by a periodic DFT calculation for the perfect
reference system. The summation of $\mu$ includes all the ions which
have contribution to the charge density in the unit cell.

\begin{figure}
\begin{center}
\includegraphics[width=\textwidth]{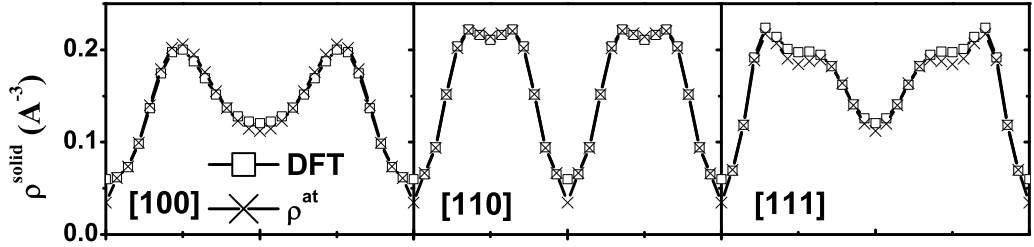}
\caption{The solid charge density of the perfect Al lattice along
[100], [110] and [111] directions obtained by the periodic DFT
calculation and the superposition of the fitted $\rho^{\rm at}$.}
 \label{rhoatom}
\end{center}
\end{figure}

In this paper, we employ the parameterized multiple Slater-type
orbitals (MSTO) \citep{Clementi1974} for the expansion of $\rho^{\rm
at}(r)$. With MSTO, the atomic wave function $\Phi$ of
many-electronis is the superposition of all relevant atomic
orbitals: $\Phi(r,\theta,\varphi)=\sum_i c_i
\phi_i(r,\theta,\varphi)$, where $c_i$ is the weight of orbital $i$
in the expansion and the $i$th atomic orbital can be written as
\begin{equation}
\label{equ:sec2:025}
\begin{split}
\phi_i(r,\theta,\varphi)=Ar^{n-1}e^{-\zeta
r}Y_{l}^m(\theta,\varphi),
\end{split}
\end{equation}
where $n$, $l$, $m$ are the principal, angular momentum and magnetic
quantum number of the orbital. $Y_{l}^m(\theta,\varphi)$ is
spherical harmonic function and $\zeta$ is related to the effective
charge of the ion. $A$ is a normalization constant and is expressed
as $A=(2\zeta)^{n+1/2}/\sqrt{(2n)!}$. With this expansion, the
atomic-centered charge density can be calculated as
\begin{equation}
\label{equ:sec2:026} \rho^{\rm
at}(r)=\int_{-\pi}^{\pi}\int_0^{2\pi}|\Phi^\ast(r,\theta,\varphi)
\Phi(r,\theta,\varphi)| d\theta d\varphi.
\end{equation}
The parameters $c_i$ and $\zeta$ are determined by minimizing Eq.
(\ref{equ:sec2:024}) with the constraint of preserving the correct
number of valence electrons. In Fig. \ref{rhoatom}, we present the
solid charge density $\rho^{\rm solid}$ determined from $\rho^{\rm
at}$ and from the periodic DFT calculation for a perfect Al lattice.
It can be seen that the constructed $\rho^{\rm solid}$ reproduces
very well the solid charge density calculated by DFT calculations
for the same perfect lattice.

\section{Computational details}
\subsection{Model setup}
\begin{figure}[htp]
\centering
\includegraphics[width=0.8\textwidth]{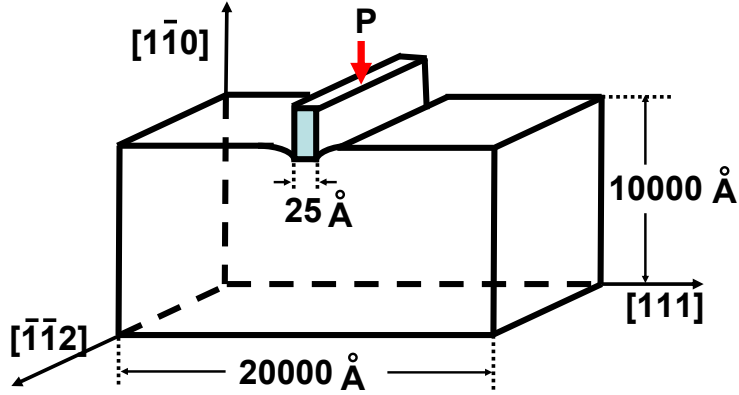}
 \caption{\label{fig:1} Schematic representation of the nanoindentation of Al thin film showing the relevant dimensions and orientations. }
\end{figure}
The present QCDFT approach is applied to nanoindentation of an Al
thin film resting on a rigid substrate with a rigid knife-like
indenter. The QCDFT method is appropriate for the problem because it
allows the modeling of system dimensions on the order of microns and
thus minimizes the possibility of contaminating the results by the
boundary conditions arising from small model sizes typically used in
MD simulations. The reason we chose this particular system is
because there exists a good kinetic energy functional and an
excellent EAM potential \citep{eam} for Al. In this paper, we have
rescaled the ``force-matching" EAM potential of Al \citep{eam} so
that it matches precisely the DFT value of the lattice constant and
bulk modulus of Al \citep{Peng2008}.

The crystallographic orientation of the system is displayed in
\fig{fig:1}. The size of the entire system is 2 $\mu$m $\times$ 1
$\mu$m $\times$ 4.9385 \AAA along the [111] (x direction), the
[$\bar{1}$10] (y direction), and the [$\bar{1}\bar{1}2$] (z
direction), respectively. The system is periodic in the z-dimension
and has the Dirichlet boundary conditions in the other two
directions. The entire system contains over 60 million Al atoms - a
size that is well beyond the reach of any full-blown quantum
mechanical calculation. The unloaded system is a perfect single
crystal similar to the experimental situation. The film is oriented
so that the preferred slip system $\left< 110 \right> \{ 111 \} $ is
parallel to the indentation direction to facilitate dislocation
nucleation. The indenter is a rigid flat punch of width 25 \AA. We
assume the perfect-stick boundary condition for the indenter so that
the Al atoms in contact with it are not allowed to slip. The
knife-like geometry of the indenter is dictated by the pseudo
two-dimensional (2D) nature of the QC model adopted.
Three-dimensional QC models do exist and can be implemented in QCDFT
\citep{Knap2003, LQC2005,LQC2006}. We chose to work with the
pseudo-2D model in this example for its simplicity. The prefix {\it
pseudo} is meant to emphasize that although the analysis is carried
out in a 2D coordinate system, the out-of-plane displacements are
allowed and all atomistic calculations are three-dimensional. Within
this setting only dislocations with line directions perpendicular to
the xy plane can be nucleated.

The simulation is performed quasistatically with a displacement
control where the indentation depth ($d$) is increased by 0.1 \AAA
at each loading step. Because DFT calculations are much more
expensive than EAM, we use EAM-based QC\citep{Tadmor1999} to relax
the system for most of the loading steps first. At $d=2.0, 3.0,
5.5,6.0,7.0,7.1,7.5$ \AA, the corresponding EAM configurations are
further relaxed by QCDFT. The QCDFT loadings are carried out after
$d=7.5$ \AA\ starting from the full relaxed EAM-QC configuration of
previous loading step, until the onset of the plasticity occurs at
$d=8.2$ \AA. Such a simulation strategy is justified based on two
considerations: (1) an earlier nanoindentation study of the same Al
surface found that the onset of plasticity occurred at a smaller
load with EAM-based local QC calculations comparing to OFDFT-based
local QC calculations \citep{LQC2005}. The result was obtained by a
local elastic stability analysis with EAM and OFDFT calculations of
energetics and stress. This suggests that we will not miss the onset
of plasticity with the present loading procedure by performing
EAM-QC relaxations preceding QCDFT. (2) Before the onset of
plasticity, the load-displacement response is essentially linear
with the slope determined by the elastic properties of the material.
In other words, only two QCDFT calculations are required to obtain
the linear part of the loading curve.

\begin{figure}[htp]
\centering
\includegraphics[width=0.8\textwidth]{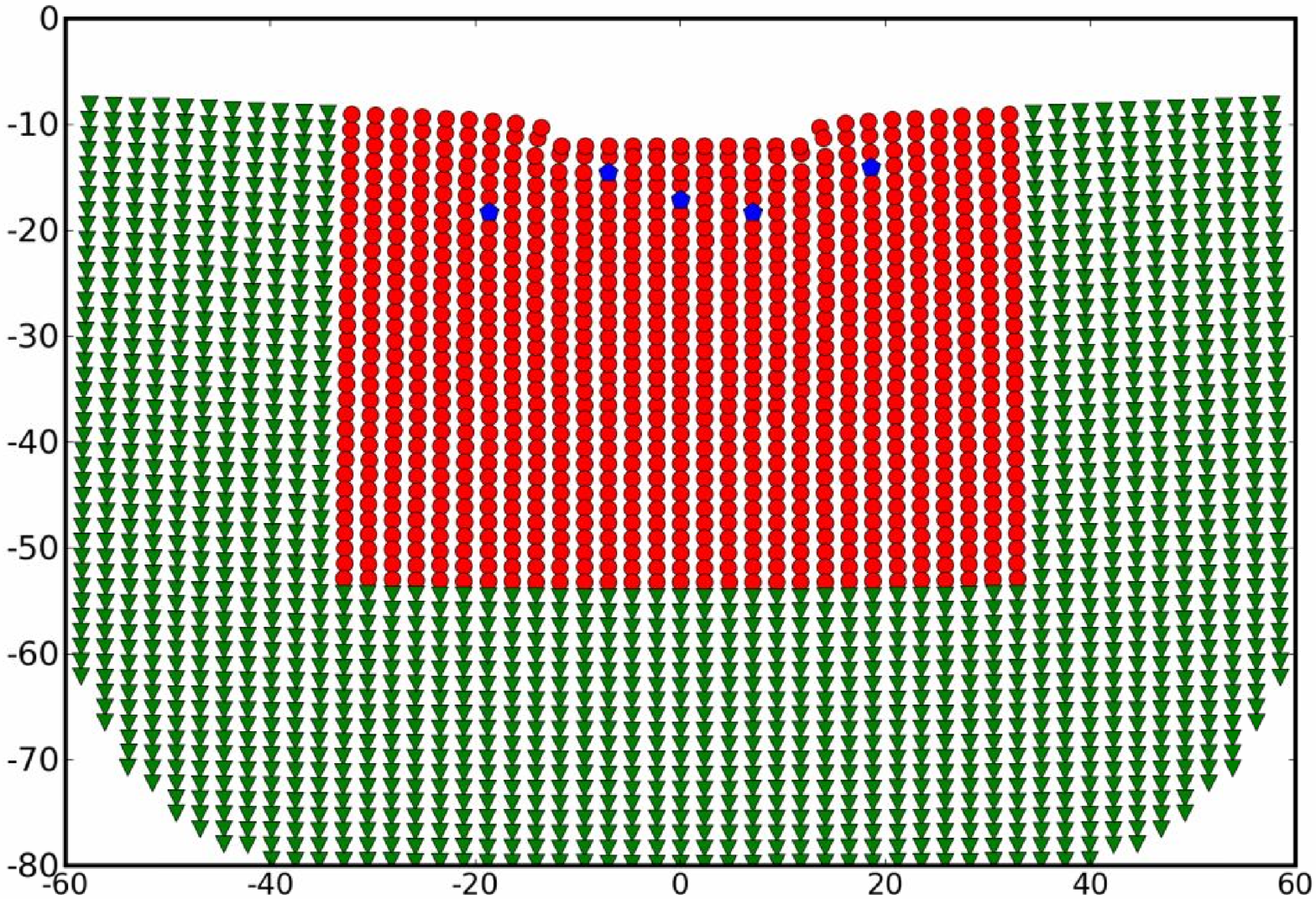}
 \caption{\label{fig:random} Schematic diagram of the randomly distributed Mg
impurities in the Al thin film. The red spheres and blue pentagons
represent nonlocal Al and Mg atoms, respectively. The green triangle
represents Al buffer atoms. The dimensions are given in \AAA.}
\end{figure}

We also study the effect of Mg impurities on the ideal strength and
incipient plasticity of the Al thin film. In the calculations, five
Mg impurities are introduced randomly below the indenter, as
schematically shown in \fig{fig:random}. The results of the randomly
distributed Mg impurities are referred as {\em random},
distinguishing from the results of the pure system, referred as {\em
pure}. At $d=3.0, 6.0, 7.5$ \AA, the {\em random} results are
obtained after full relaxations of the {\em pure} Al system. The
QCDFT loading is carried out after $d=7.5$ \AA\ starting from the
full relaxed configuration of a previous loading step, until the
onset of the plasticity occurs at $d=8.1$ \AA.

The parameters of the OFDFT density-dependent kernel are chosen from
reference \citep{Wang1999}, and Al ions are represented by the
Goodwin-Needs-Heine local pseudopotential \citep{Goodwin1990}. The
high kinetic energy cutoff for the plane wave basis of 1600 eV is
used to ensure the convergence of the charge density. For the
nonlocal calculation, the grid density for the volume $\Omega^{\rm
I}$ is 5 gridpoints per \AA. The $\Omega^{\rm I}$ box goes beyond
the nonlocal region by 8 \AAA in $\pm$x and $\pm$y directions so
that $\rho^{\rm I}$ decays to zero at the boundary of $\Omega^{\rm
I}$, as shown in \fig{fig:omega}. The relaxation of all repatoms is
performed by a conjugate gradient method until the maximum force on
any repatom is less than 0.03 eV/\AA.

\section{Results and Analysis}
The load-displacement curve is the typical observable for
nanoindentation, and is widely used in both experiment and theory,
often serving as a link between the two. In particular, it is
conventional to identify the onset of incipient plasticity with the
first drop in the load-displacement curve during indentation
\citep{Corcoran1997,Suresh1999,Gouldstone2000,Tadmor1999,Shenoy2000,
Zhu2004,Knap2003,LQC2005,Gouldstone2007,Peng2008}. In the present
work, the load is given in N/m, normalized by the length of the
indenter in the out-of-plane direction.

\begin{figure}[htp]
\centering
\includegraphics[width=0.75\textwidth]{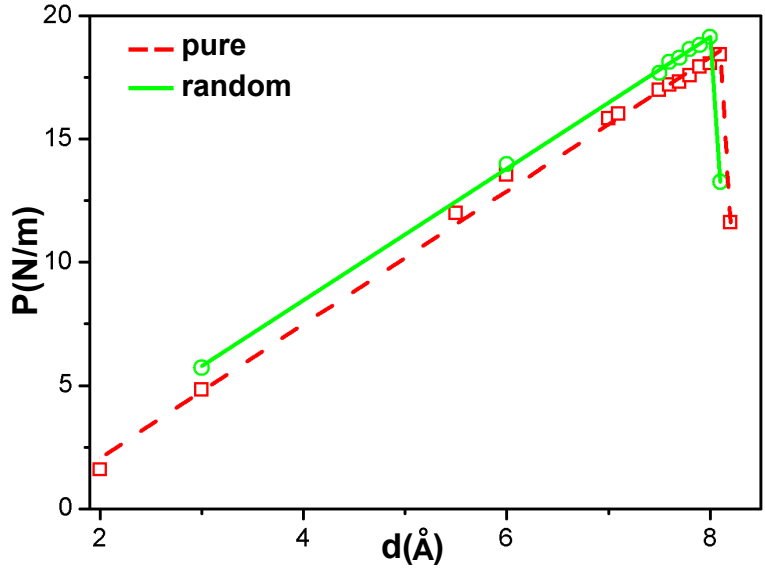}
 \caption{\label{fig:ph} Load-displacement plot for the nanoindentation
of the Al thin film with a rigid rectangular indenter: pure Al (red
squares) and randomly distributed Mg impurity system (green
circles). The corresponding lines are the best fit to the data
points. }
\end{figure}

\begin{figure}[htp]
\includegraphics[width=\ww\textwidth]{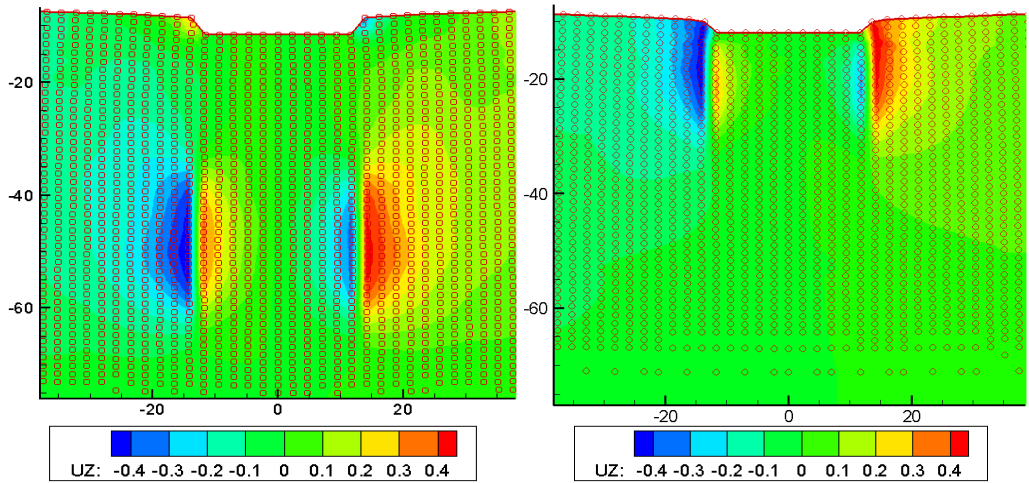}
 \caption{\label{fig:DFT} The out-of-plane displacement $u_z$ obtained from the pure (left) and
with Mg impurities (right) QCDFT calculations. The circles represent
the repatoms and the displacement ranges from -0.4 (blue) to 0.4
(red) \AA.}
\end{figure}

For pure Al, the load-displacement ($P-d$) curve shows a linear
relation followed by a drop at $d = 8.2$ \AA, shown by the dashed
line in \fig{fig:ph}. The drop corresponds to the homogeneous
nucleation of dislocations beneath the indenter - the onset of
plasticity. A pair of straight edge dislocations are nucleated at
x=$\pm$13 \AA, and y=-50 \AA. In \fig{fig:DFT}, we present the
out-of-plane (or screw) displacement $u_z$ of the nonlocal repatoms.
The non-zero screw displacement of the edge dislocations suggests
that each dislocation is dissociated into two 1/6 $<$112$>$ Shockley
partials bound by a stacking fault with a width of about 19 \AA. The
activated slip planes are those \{111\} planes that are adjacent to
the edges of the indenter. The slope for the linear part of the
curve is 27.1 GPa, which is less than the shear modulus $\mu$=33.0
GPa and $C_{44}=29.8$ GPa. The critical load, $P_{cr}$ for the
homogeneous dislocation nucleation is 18.4 N/m, corresponding to a
hardness of 7.3 GPa (the critical load normalized by the area of the
indenter), which is 0.22 $\mu$ . The drop in applied load due to the
nucleation of dislocations is $\Delta P= 6.8$ N/m, agreeing with the
load drop estimated by the elastic model\citep{Tadmor1999} which is
$\Delta P= 7.7$ N/m.

For randomly distributed impurities in the Al thin film, the
load-displacement curve shows a linear relation up to a depth of 8.0
\AA, followed by a drop at $d = 8.1$ \AA, as shown by the solid line
in \fig{fig:ph}. The slope of initial linear part of the
load-displacement curve is $26.7$ GPa, rather close to the
corresponding pure Al value. The maximum load in linear region is
$P^{im}_{cr}=19.2$ N/m, corresponding to a hardness of 7.6 GPa,
which is 0.3 GPa greater than the pure Al system. A pair of Shockley
partial dislocations is nucleated at x=-13 \AAA, y=-25 \AAA and x=13
\AAA, y=-22 \AAA respectively as shown in the right panel of
\fig{fig:DFT}. The drop in the applied load due to the dislocation
nucleation is $5.9$ N/m. The estimated load drop by the elastic
model is $\Delta P= 7.6$ N/m. The smaller drop of the load for the
random case than the elastic model is probably due to the presence
of the Mg impurities, which is not accounted for in the elastic
model \citep{Tadmor1999}. The fact that the critical load and the
hardness of the Al-Mg alloy are greater than that of the pure Al
system demonstrates that the Mg impurities are responsible for the
solid solution strengthening of the Al thin film. The presence of Mg
impurities also hinders the formation of full edge dislocations and
as a result, only partial dislocations are nucleated and they are
pinned near the surface as shown in \fig{fig:DFT}.

Finally we point out the possibility that the emitted dislocations
may be somewhat constrained by the local/nonlocal interface from
going further into the bulk. Because the critical stress to move an
edge dislocation in Al is vanishingly small (~$10^{-5}\mu$)
comparing to that to nucleate a dislocation ($10^{-1}\mu$), a small
numerical error in stress could easily lead to a large difference in
the equilibrium dislocation position. The four-order-of-magnitude
disparity poses a significant challenge to all atomistic simulations
in predicting dislocation nucleation site, QCDFT method included.
One can only hope to obtain a reliable critical load for the
incipient plasticity, rather than for the equilibrium position of
dislocations. The same problem has been observed and discussed by
others \citep{qc2}. However, despite the problem, the dramatic
difference observed in the two panels of Fig. 7 unambiguously
demonstrates the strengthening effect of Mg impurities. Therefore
the conclusion is still valid.

\section{CONCLUSION}

In summary, we propose a concurrent multiscale method that makes it
possible to simulate multi-million atoms based on the density
functional theory. The method - QCDFT - is formulated within the
framework of the QC method, with DFT as its sole energy input. The
full-blown DFT and DFT-based elasticity theory would be the two
limiting cases corresponding to a fully nonlocal or a fully local
version of QCDFT. The QCDFT method is applied to nanoindentation of
an Al thin film in the presence and absence of randomly distributed
Mg impurities. The Mg impurities are found to strengthen the
hardness of Al and hinder the dislocation nucleation. The results
suggest that QCDFT is a promising method for quantum simulation of
materials properties at length scales relevant to experiments.

\section{ACKNOWLEDGEMENTS} The work at California State University Northridge was supported by
NSF PREM grant DMR-0611562 and DoE SciDAC grant DE-FC02-06ER25791.

\bibliographystyle{elsarticle-harv}
\bibliography{CMS}

\begin{thebibliography}{25}
\bibitem[{Lin and Truhlar(2007)}]{Lin2007}
Lin, H., Truhlar, D.~G., 2007. {QM/MM}: What have we learned, where are we, and
  where do we go from here? Theor. Chem. Acc. 117, 185.
\end{thebibliography}

\end{document}